\documentclass[a4paper]{article}
\usepackage[utf8]{inputenc}
\usepackage[english]{babel}
\usepackage{amsmath}
\usepackage{authblk}
\usepackage{graphicx}
\usepackage{units}
\usepackage[top=2.5cm,left=2.5cm,right=2.5cm,bottom=2.5cm]{geometry}
\usepackage{url}

\def\l{\left}
\def\r{\right}
\def\inv{\mathrm{inv}}
\def\LCMS{\mathrm{LCMS}}
\def\ev{\mathrm{ev}}
\def\bin{\mathrm{bin}}

\title{The Shape of the Correlation Function}

\author[1]{Jakub Cimerman}
\author[2]{Boris Tom\' a\v sik}
\author[3]{Christopher Plumberg}
\affil[1]{FNSPE, \v Cesk\' e vysok\' e u\v cen\' i technick\' e v Praze, B\v rehov\'a 7, 11519 Praha 1, Czechia; jakub.cimerman@fjfi.cvut.cz}
\affil[2]{Univerzita Mateja Bela, Tajovsk\' eho 40, 97401 Bansk\' a Bystrica, Slovakia\newline 
and FNSPE, \v Cesk\' e vysok\' e u\v cen\' i technick\' e v Praze, B\v rehov\'a 7, 11519 Praha 1, Czechia; boris.tomasik@cern.ch}
\affil[3]{Department of Astronomy and Theoretical Physics, Lund University, S\"olvegatan 14A, SE-223 62, Lund,
Sweden; christopher.plumberg@thep.lu.se}

\date{}

\begin{document}

\maketitle

\begin{abstract}
The correlation function measured in ultrarelativistic nuclear collisions is non-Gaussian. By making use of models we discuss and assess how much various effects can influence its shape. In particular, we focus on the parametrisations expressed with the help of L\'evy-stable distributions. We show that the L\'evy index may deviate substantially from 2 due to non-critical effects such as non-spherical shape, resonance decays, event-by-event fluctuations and functional dependence on $Q_{\inv}$ or similar.

\end{abstract}

\section{Introduction}

Correlation femtoscopy \cite{cf1, cf2} has become a standard technique for the experimental analysis of heavy-ion collisions. Usually, the two-particle correlation functions are fitted to a Gaussian form. However, the real shape of the correlation function is often strongly non-Gaussian and is often better described by a L\'evy-stable distribution. A L\'evy index much below 2 has recently been observed experimentally \cite{levy}.
%There are papers which obtained the value of the L\'evy index below 1 \cite{levy}. 
It has been suggested that even lower value of the L\'evy index equal to 0.5 may identify matter produced at the critical endpoint of the QCD phase diagram. Despite this, there are some non-critical effects which can affect the value of the L\'evy index significantly.

%%%%%%%%%%%%%%%%%%%%%%%%%%%%%%%%%%%%%%%%%%%%%%%%%%%%%%%%%%%%%%%%%%%%%%%%%%%

\section{HBT Formalism}

The two-particle correlation function probes the momentum-space structure of correlations between pairs of particles produced in heavy-ion collisions. In this work, we focus on correlations between charged pion pairs. The correlation function is constructed as the ratio of the two-particle spectrum to the product of two one-particle spectra, evaluated at momenta $p_1$ and $p_2$:
\begin{equation}
C(p_1, p_2) = \dfrac{P_2(p_1, p_2)}{P(p_1)P(p_2)} = \frac{E_1E_2\frac{\mathrm{d}^6N}{\mathrm{d}p_1^3\mathrm{d}p_2^3}}{\left(  E_1 \frac{\mathrm{d}^3N}{\mathrm{d}p_1^3}\right)\left( E_2\frac{\mathrm{d}^3N}{\mathrm{d}p_2^3} \right)}.
\end{equation}
The correlation function is often expressed in terms of the momentum difference and the average momentum
\begin{equation}
q=p_1-p_2, \qquad K=\frac{1}{2}(p_1+p_2).
\end{equation}

The source of particle production can be described by the emission function $S(x,p)$ which describes the probability that a particle with momentum $p$ is emitted from position $x$. The particle spectrum can then be calculated by integration of the emission function over the fireball volume. 

Due to symmetrisation of the wave function of pairs of bosons, there is a peak in the correlation function for small $q$. Since we are interested in studying the region of the peak itself, we use the smoothness approximation, where $K\approx p_1 \approx p_2$. Using this approximation the correlation function takes the form
\begin{equation}
C(q,K)-1\approx\dfrac{\l|\int \mathrm{d}^4xS(x,K) e^{iqx} \r|^2}{\left(\int \mathrm{d}^4xS(x,K)\right)^2}.
\end{equation}

The Gaussian parametrisation of the correlation function reads
%The information about the homogeneity region can be then obtained by parametrizing the correlation function by Gaussian form
\begin{equation}
C_G(\vec{q},\vec{K}) = 1+ \lambda (\vec{K}) \exp \left[ - \sum_{i,j=o,s,l} R_{ij}^2(\vec{K}) q_i q_j\right].
\label{eq-gauss}
\end{equation}
Here, the HBT radii (in the out-side-long system) $R_{ij}^2 (\vec{K})$ can be understood as lengthscales characterizing the homogeneity region which produces pion pairs with average momentum $K$, and $\lambda$ quantifies the magnitude of the correlation function when $\vec{q}=0$. 

Nevertheless, since the Gaussian parametrisation often does not adequately describe the experimentally measured correlation function, we also use L\'evy parametrisation of the correlation function 
\begin{equation}
C_L(\vec{q},\vec{K}) = 1+ \lambda'(\vec{K}) \exp \left[ - \Bigg| \sum_{i,j=o,s,l} R_{ij}^{'2}(\vec{K}) q_i q_j \Bigg| ^{\alpha/2} \right].
\label{eq-3dlevy}
\end{equation}
The parameters $\lambda'$ and $R_{ij}^{'2}$ are analogous to those used in the Gaussian parametrisation, but their values may differ from their Gaussian counterparts, and they also have no direct correspondence with the source widths often used to estimate and interpret the Gaussian $R^2_{ij}$. The additional parameter $\alpha$ is known as the L\'evy index and controls the form of the distribution used to approximate correlation function: for  $\alpha=2$ L\'evy distribution becomes  Gaussian distribution, while for  $\alpha=1$ it becomes exponential distribution.

When we use one-dimensional projection of relative momentum $q$, we also use a corresponding one-dimensional L\'evy parametrisation
\begin{equation}
C_L(Q) = 1 + \lambda' \exp (-|R'Q|^\alpha).
\label{eq-1dlevy}
\end{equation}

%%%%%%%%%%%%%%%%%%%%%%%%%%%%%%%%%%%%%%%%%%%%%%%%%%%%%%%%%%%%%%%%%%%%%%%%%%%

\section{Effects Leading to Non-Gaussianities}

There are four effects we studied which can lead to non-Gaussianities. The first is  event averaging. Each event possesses a variety of properties $-$ such as size, geometric and dynamical anisotropies, and so on $-$ which tend to fluctuate randomly from one event to the next. In order to build up statistics it is conventional to average correlation function over a large number of different events. The formula for the  correlation function thus must be replaced by
\begin{equation}
C(q,K)\approx 1+\dfrac{\left\langle \l|\int \mathrm{d}^4xS(x,K)e^{iqx}\r|^2\right\rangle_{\ev}}{\left\langle\left(\int \mathrm{d}^4xS(x,K)\right)^2\right\rangle_{\ev}}.
\end{equation}

Another way to improve statistical precision is to use a one-dimensional projection of the relative momentum. The correlation function is then a function of a single scalar quantity. There are two ways to perform this projection: either using Lorentz-invariant variable
\begin{equation}
%Q_{LI}^2 \equiv 
Q^2_{\inv}= -q^\mu q_\mu = \vec{q}\cdot \vec{q}-(q^0)^2
\label{eq-qli}
\end{equation}
or a longitudinally boost-invariant one \cite{levy}
\begin{equation}
Q_{\LCMS}^2=\sqrt{(p_{1x}-p_{2x})^2+(p_{1y}-p_{2y})^2+q_{\mathrm{long},\LCMS}^2},
\label{eq-qlbi}
\end{equation}
where $q_{\mathrm{long},\LCMS}^2=\frac{(p_{1z}E_2-p_{2z}E_1)^2}{K_0^2-K_l^2}$.

The next effect which can influence the shape of the correlation function is averaging with respect to the pair momentum $\vec{K}$.
When measuring the  correlation function  bins in $\vec{K}$ must be created which cannot be taken arbitrarily small. Thus the correlation function is averaged over some pair momentum interval which leads to an adjustment in the formula for the correlation function
\begin{equation}
C(q,K)\approx 1+\dfrac{\int_{\bin}\mathrm{d}^3K\l|\int \mathrm{d}^4xS(x,K)e^{iqx}\r|^2}{\int_{\bin}\mathrm{d}^3 K\left(\int \mathrm{d}^4xS(x,K)\right)^2}.
\end{equation}

The last effect we studied is the impact of resonance decays on the L\'evy index. Different resonances contribute to the correlation function with different lengthscales and timescales, while the Gaussian function is given by only a single lengthscale. Therefore, the correlation function must deviate from a Gaussian form once resonance effects are included.

%%%%%%%%%%%%%%%%%%%%%%%%%%%%%%%%%%%%%%%%%%%%%%%%%%%%%%%%%%%%%%%%%%%%%%%%%%%%

\section{Models}

To show that our results are not just model artifacts, we decided to use two different models. The first one is the blast-wave model \cite{bw}, which describes an expanding locally thermalised fireball. It is characterized by the emission function
\begin{equation}
S(x,p)\mathrm{d}^4x  =\frac{m_t\cosh (\eta- Y)}{(2\pi)^3}\mathrm{d}\eta \mathrm{d}x\mathrm{d}y\frac{\tau \mathrm{d}\tau}{\sqrt{2\pi}\Delta\tau} \exp \left(-\frac{(\tau-\tau_0)^2}{2\Delta\tau^2}\right) \exp\left(-\frac{E^\ast}{T}\right)\Theta \left(1-\overline{r}\right),
\end{equation}
where $\Theta \left(1-\overline{r}\right)$ is Heaviside step function, $E^\ast = p_\mu p^\mu$ is the energy in the rest frame of the fluid and $\overline{r}=\frac{r}{R(\theta)}$ is a scaled radius of fireball in transverse plane. The blast-wave model also contains two types of anisotropies. Spatial anisotropy is characterized by a Fourier series in the azimuthal dependence of the fireball radius:
\begin{equation}
R(\theta) = R_0\left(1-\sum_{n=2}^{\infty}a_n \cos\left(n(\theta-\theta_n)\right)\right).
\end{equation}
Simlarly, flow anisotropy reflects a distribution in the transverse rapidity:
\begin{equation}
\rho(\overline{r},\theta_b)=\overline{r}\rho_0 \left(1+\sum_{n=2}^{\infty}2\rho_n \cos\left(n(\theta_b-\theta_n)\right)\right).
\end{equation}

To generate events we use DRAGON \cite{dragon1, dragon2}, which is a Monte Carlo event generator based on the blast-wave model with added resonance decays. For this study we generated sets of 50,000 events with parameters set to: temperature $T=120\:\unit{MeV}$, the average transverse radius $R_0=7\:\unit{fm}$, freeze-out time $\tau_{fo}=10\:\unit{fm/}c$, the strength of the transverse expansion $\rho_0 = 0.8$, second order spatial anisotropy $a_2 \in (-0.1;0.1)$ and second order flow anisotropy $\rho_2 \in (-0.1;0.1)$. To calculate correlation functions from these events we used CRAB \cite{crab}.

The second model we used is a hydrodynamical model of the collision system using iEBE-VISHNU \cite{iebe1, iebe2}. It is a 2+1-dimensional hydrodynamic simulation with boost-invariant Israel-Stewart hydrodynamics equations and Glauber Monte-Carlo initial conditions. For this study we generated 1,000 events of $0-10\%$ Au+Au collisions at 200$A$ GeV with a freeze-out temperature $T_{fo}=120\:\unit{MeV}$ and $\eta/s=0.08$. To compute the HBT correlation function we used the HoTCoffeeh code \cite{hotcoffee}, which directly evaluates Cooper-Frye integrals over the freeze-out surface on an event-by-event basis. Thanks to that we can calculate correlation functions with negligible uncertainties.

%%%%%%%%%%%%%%%%%%%%%%%%%%%%%%%%%%%%%%%%%%%%%%%%%%%%%%%%%%%%%%%%%%%%%%%%%%%%

\section{Results}

\begin{figure}[h!]
\centering
\includegraphics[width=0.48\textwidth]{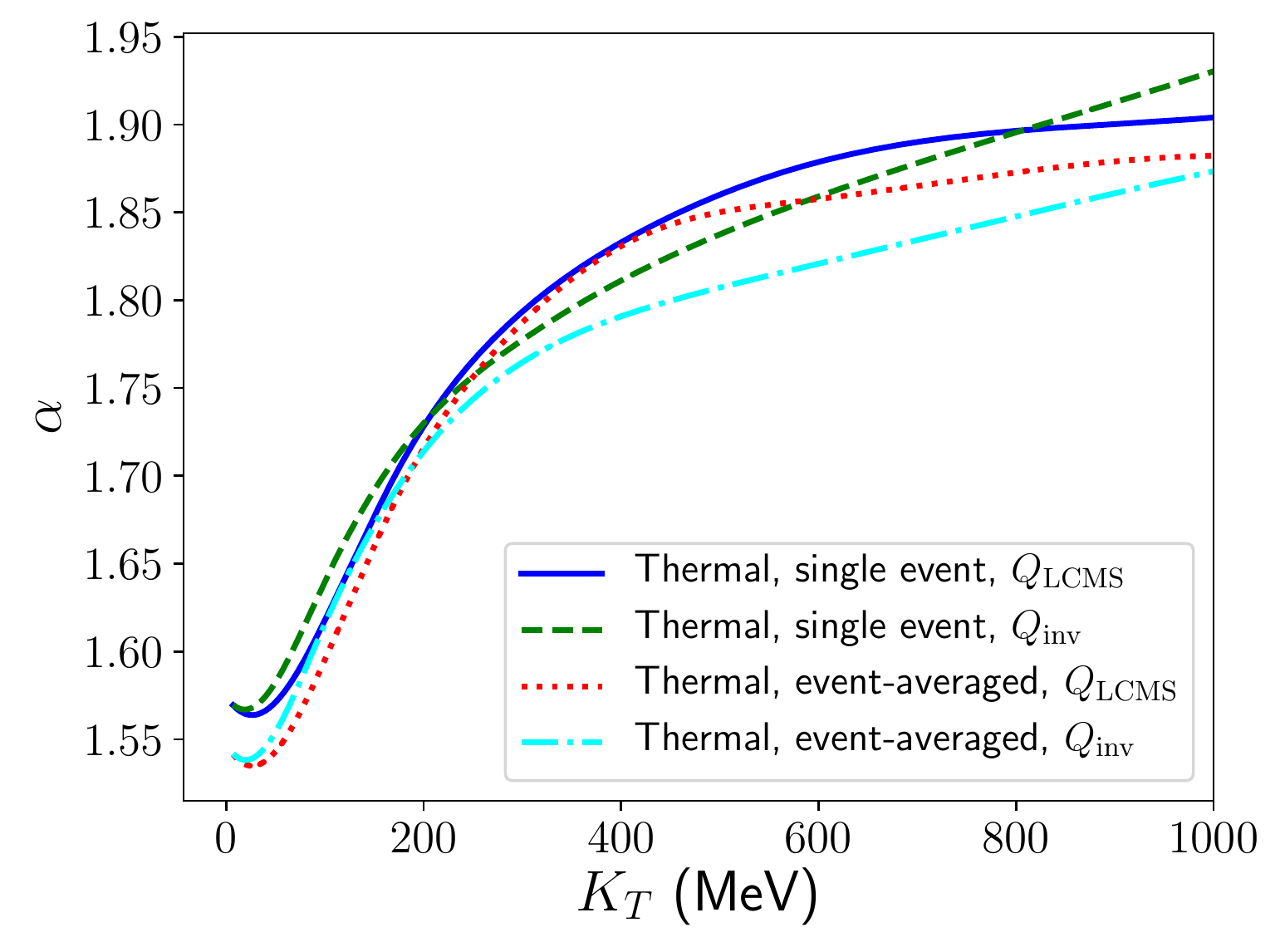}
\includegraphics[width=0.48\textwidth]{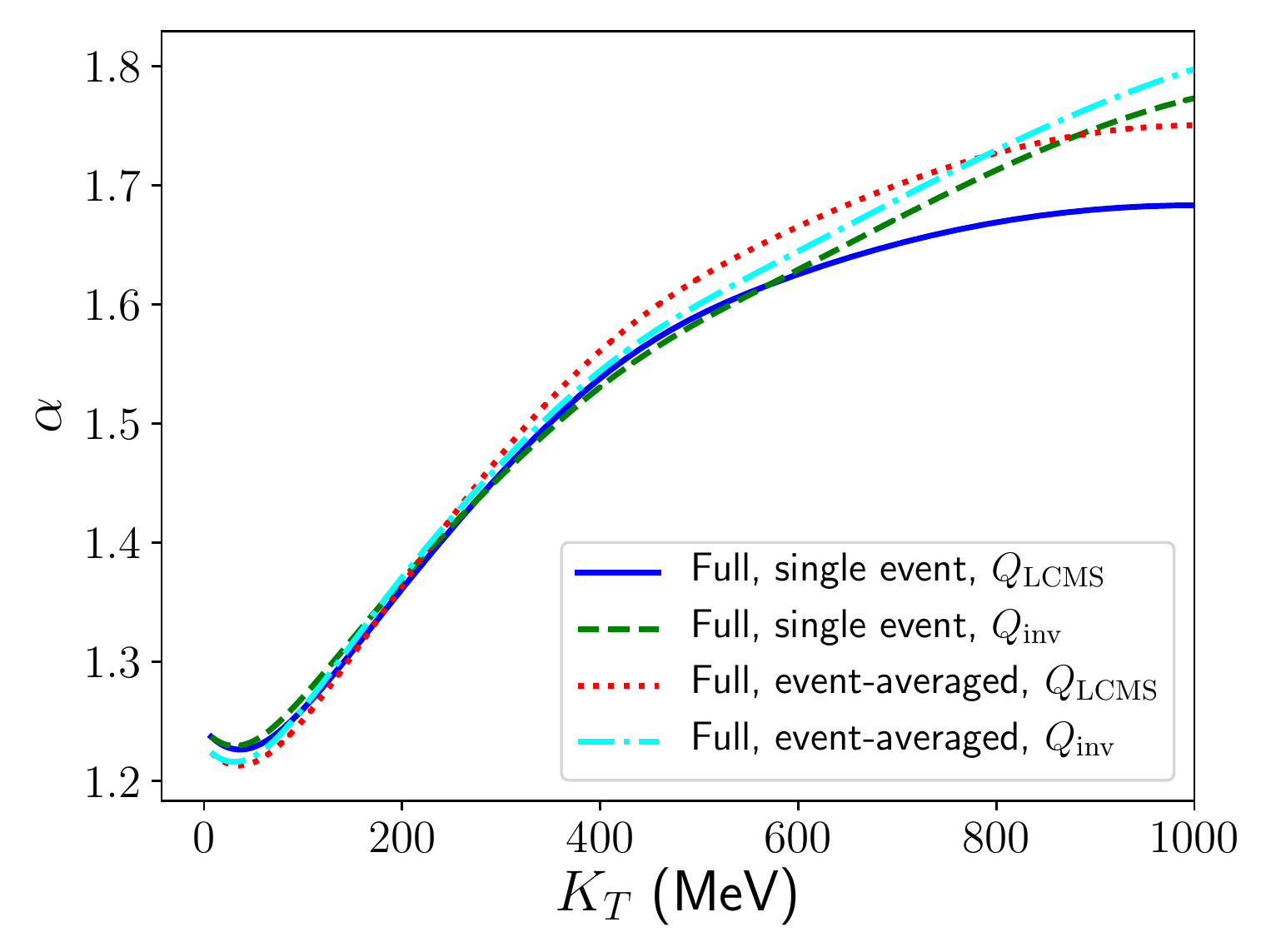}
\caption{A comparison of $\alpha (K_T)$ with and without different non-Gaussian effects in hydrodynamic model: with and without event averaging and for different choices of $Q$ (solid blue and dashed green vs. dotted red and dash-dotted cyan). The comparison is made both for thermal pions only (left panel) and for the full thermal and resonance contributions added together (right panel).}
\label{fig1}
\end{figure}

Once we have the correlation functions, we can obtain the L\'evy index with a 1D fit using Eq. \eqref{eq-1dlevy} or a 3D fit using Eq. \eqref{eq-3dlevy}. In this study, we focus on the $K_T$-dependence of the L\'evy index. First, we used the hydrodynamical model to check the relative importance of several of the effects discussed above. In Figure \ref{fig1} we see the impact of three of these:
\begin{itemize}
\item correlation function with resonances (right panel) vs. without resonances (left panel),
\item single event (solid blue and dashed green) vs. event-averaged (dotted red and dash-dotted cyan),
\item $Q_{\inv}$ (solid blue and dotted red) vs. $Q_{\LCMS}$ (dashed green and dash-dotted cyan).
\end{itemize}
From this figure, we can say that the latter two effects do not affect the L\'evy index significantly. For low $K_T$ they shift $\alpha$ by less than  0.05. For high $K_T$, the impact is larger, but the shift is still much smaller than the one due to resonances. We find that the inclusion of resonances reduces the value of the L\'evy index by 0.2-0.3. Nevertheless, as we show below, the largest effects are concentrated mainly at low $K_T$ and are due to the use of a one-dimensional projection of the relative momentum.

%%%%%%%%%%%%%%%%%%%%%%%
%%%%%%%%%%%%%%%%%%%%%%%

\begin{figure}[h!]
\centering
\includegraphics[width=0.32\textwidth]{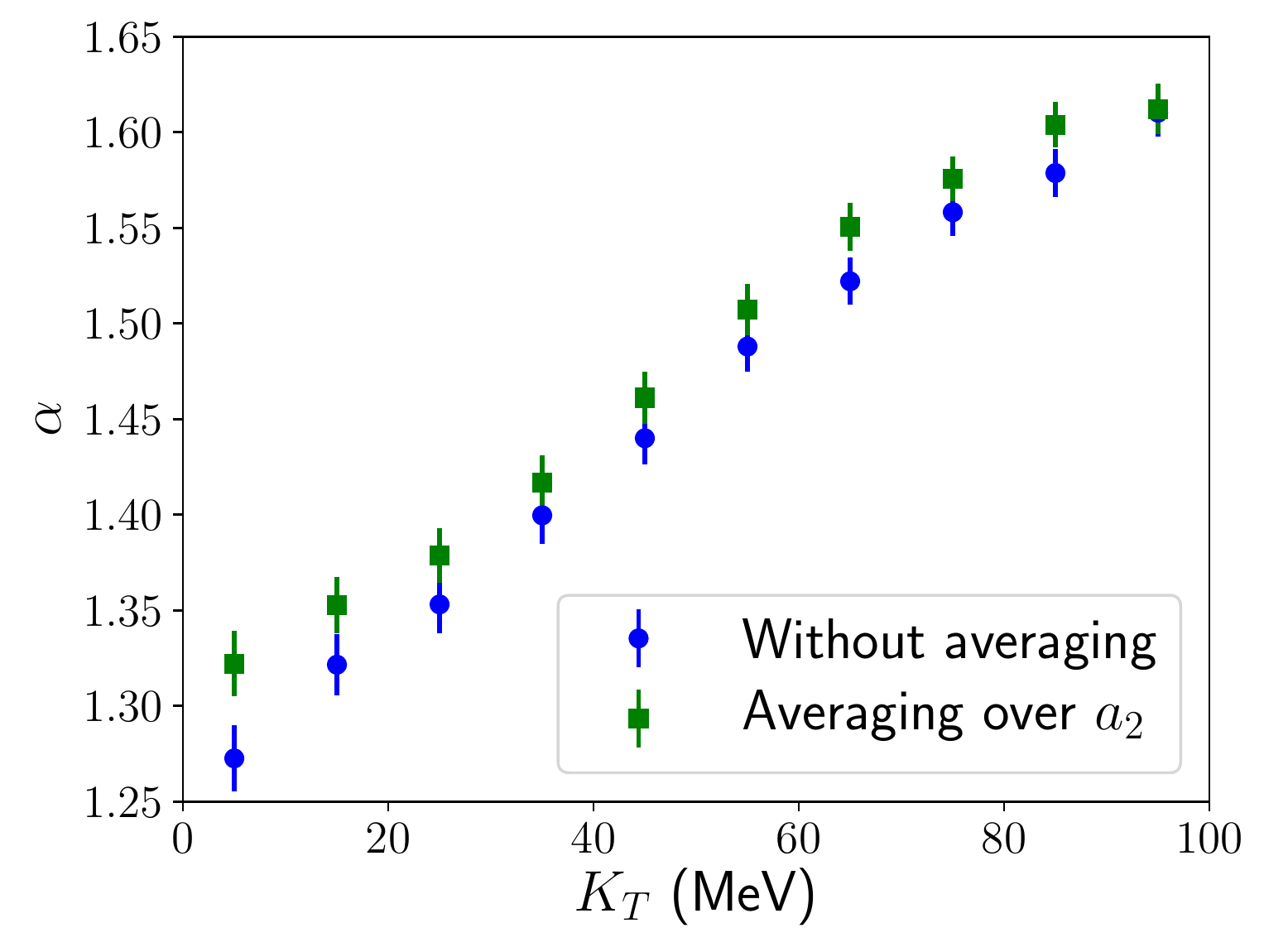}
\includegraphics[width=0.32\textwidth]{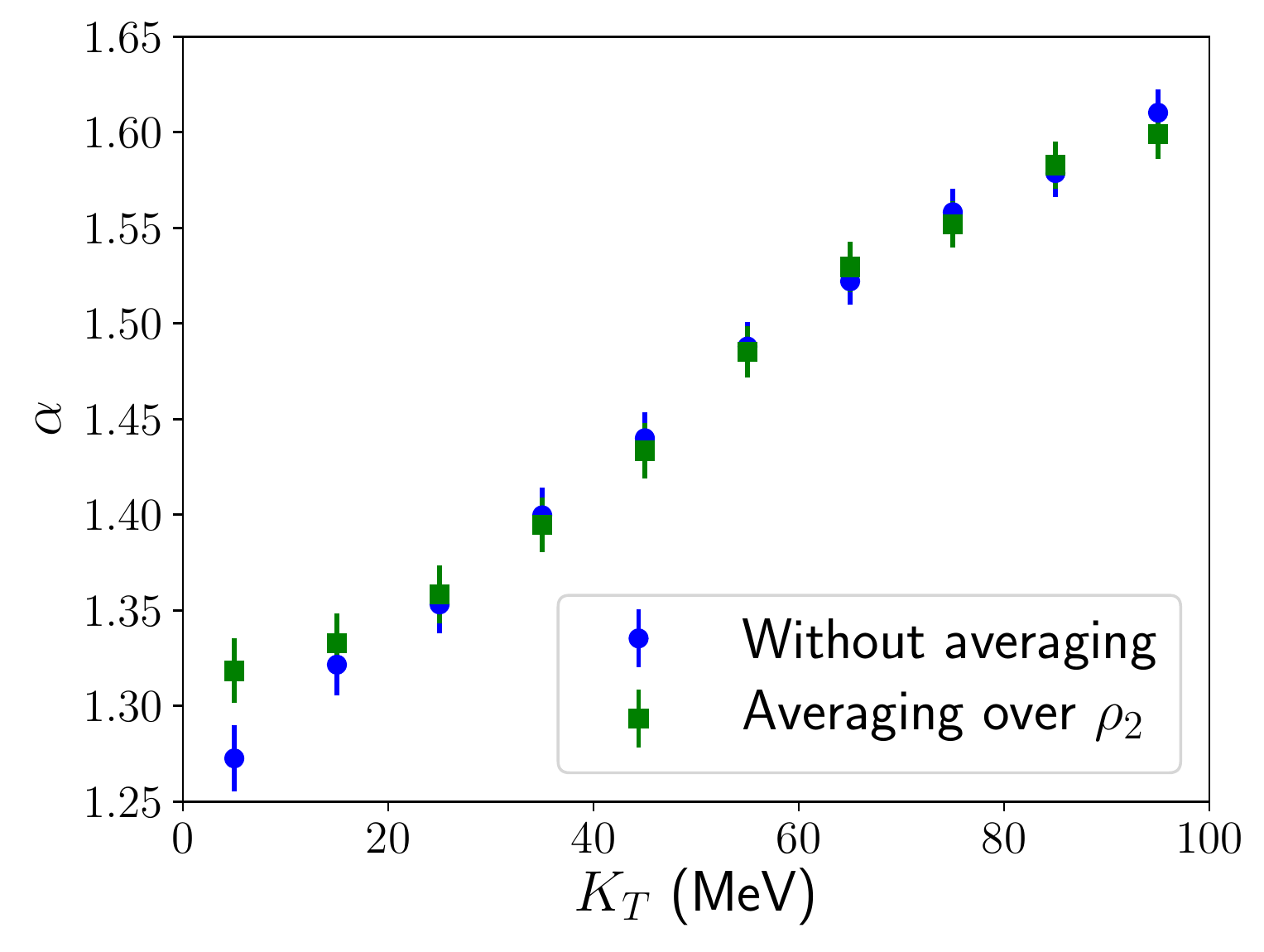}
\includegraphics[width=0.32\textwidth]{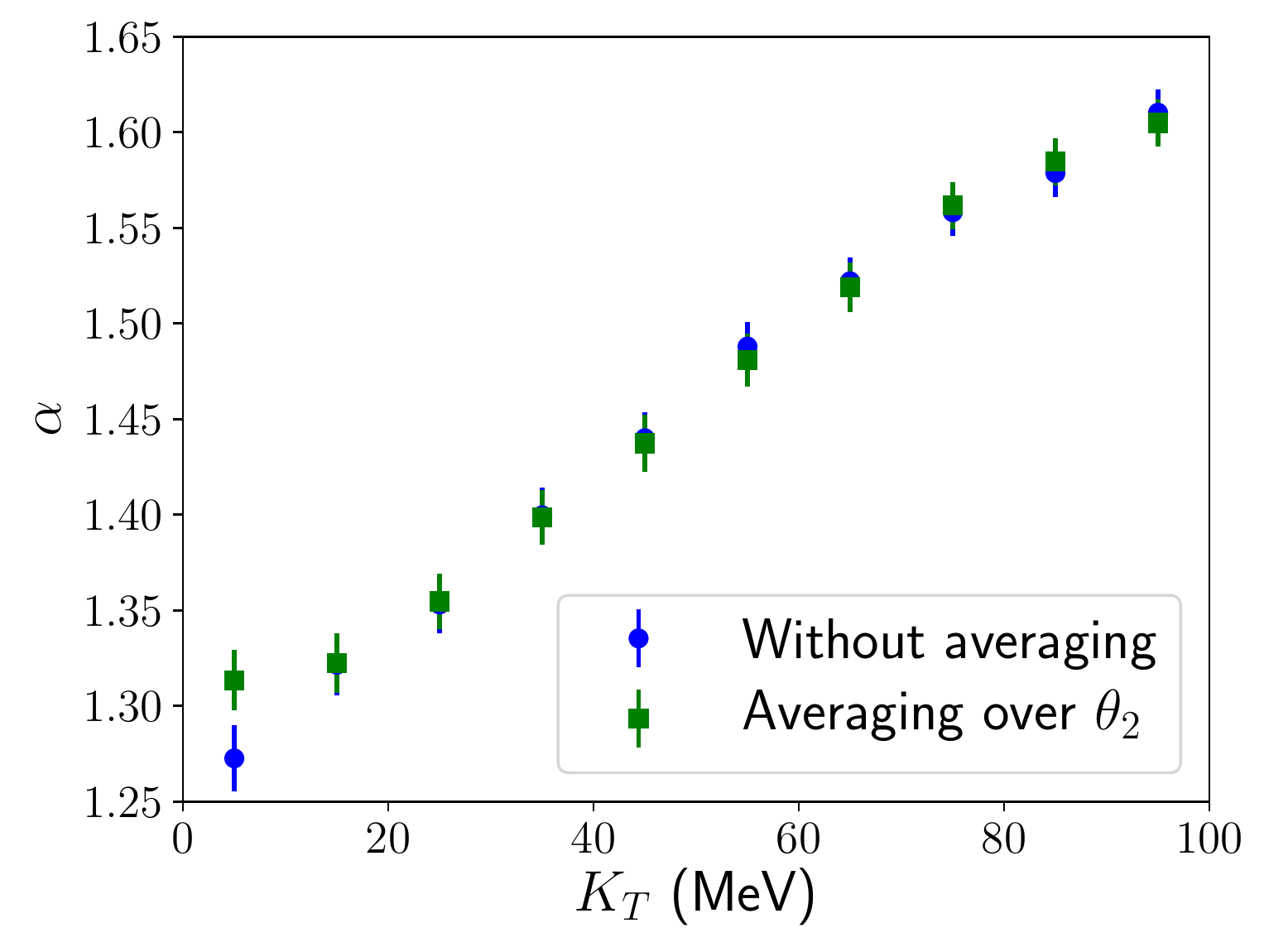}
\caption{The L\'evy index of the 1D fit to the correlation function in $Q_{\inv}$. The green points show results calculated with fixed anisotropies, while the blue points show results calculated for averaged events over $a_2$ (left), $\rho_2$ (middle) and $\theta_2$ (right).}
\label{fig2}
\end{figure}

Figure \ref{fig2} shows the influence of averaging over various parameters of the blast-wave model. Events without averaging have fixed parameters $a_2=0.05$, $\rho_2=0.05$ and $\theta_2=0$, while events with averaging have those parameters running in interval $(-0.1;0.1)$ for $a_2$ and $\rho_2$, respectively $(0;2\pi)$ for $\theta_2$. These plots show that the effect of averaging over $a_2$ is the biggest, but still smaller than the error bars. Thus we can say that this effect plays no role in the resulting value of the L\'evy index.

\begin{figure}[h!]
\centering
\includegraphics[width=0.48\textwidth]{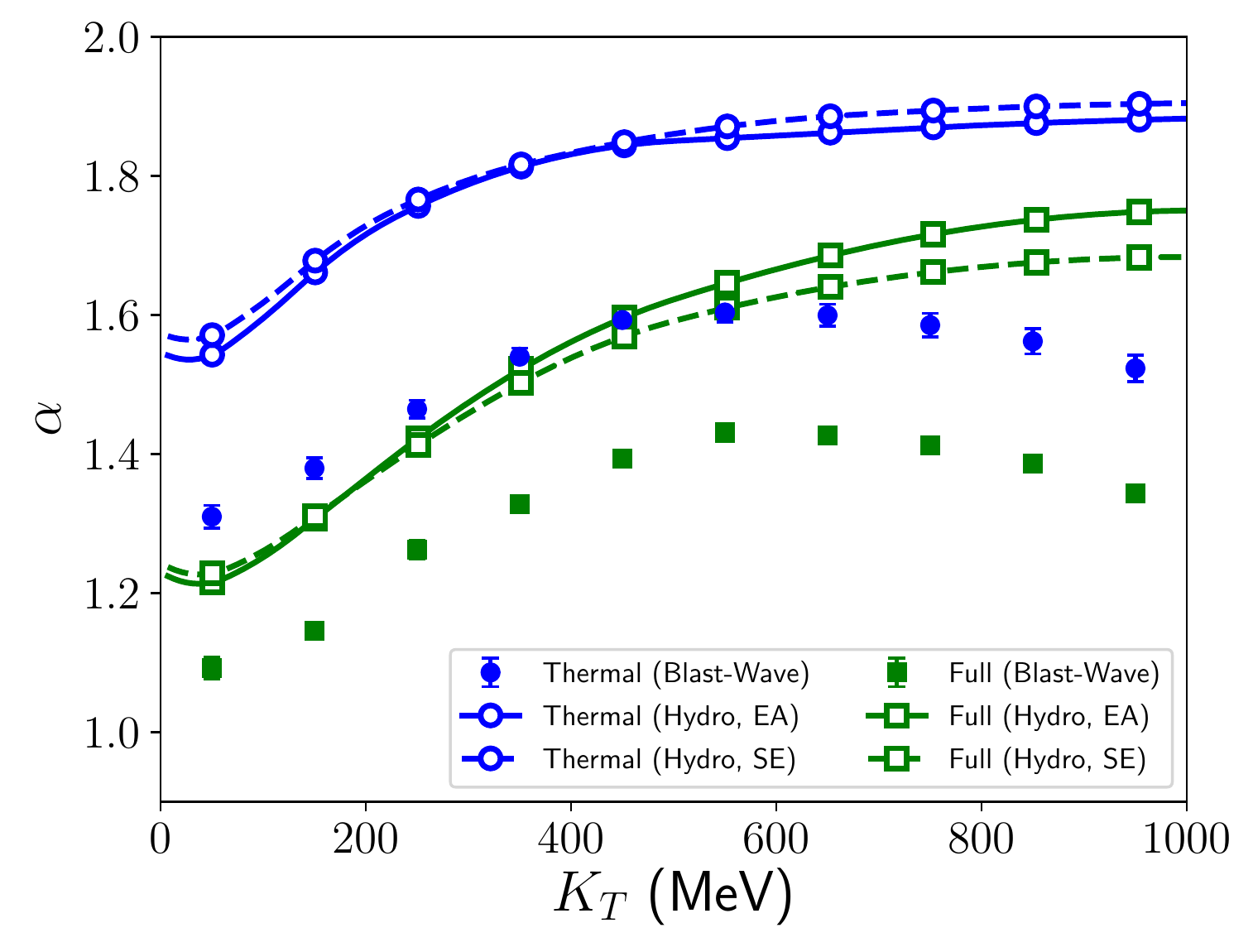}
\caption{The L\'evy index of the 1D fit to the correlation function in $Q_{\inv}$. The blue circles show results from a source without resonances, while the green squares show results from a source with resonances. The solid points with error bars correspond to the blast-wave model and the open points represent the hydrodynamic results for event-averaged (solid) and single-event (dashed) correlation functions.}
\label{fig3}
\end{figure}

To estimate the model-independent impact of resonances on the L\'evy index, we calculated its $K_T$-dependences using both our models (Figure \ref{fig3}). This plot shows that, regardless of the model used in calculations, resonances reduce the value of the L\'evy index by $\sim$0.2.

\begin{figure}[h]
\centering
\includegraphics[width=0.48\textwidth]{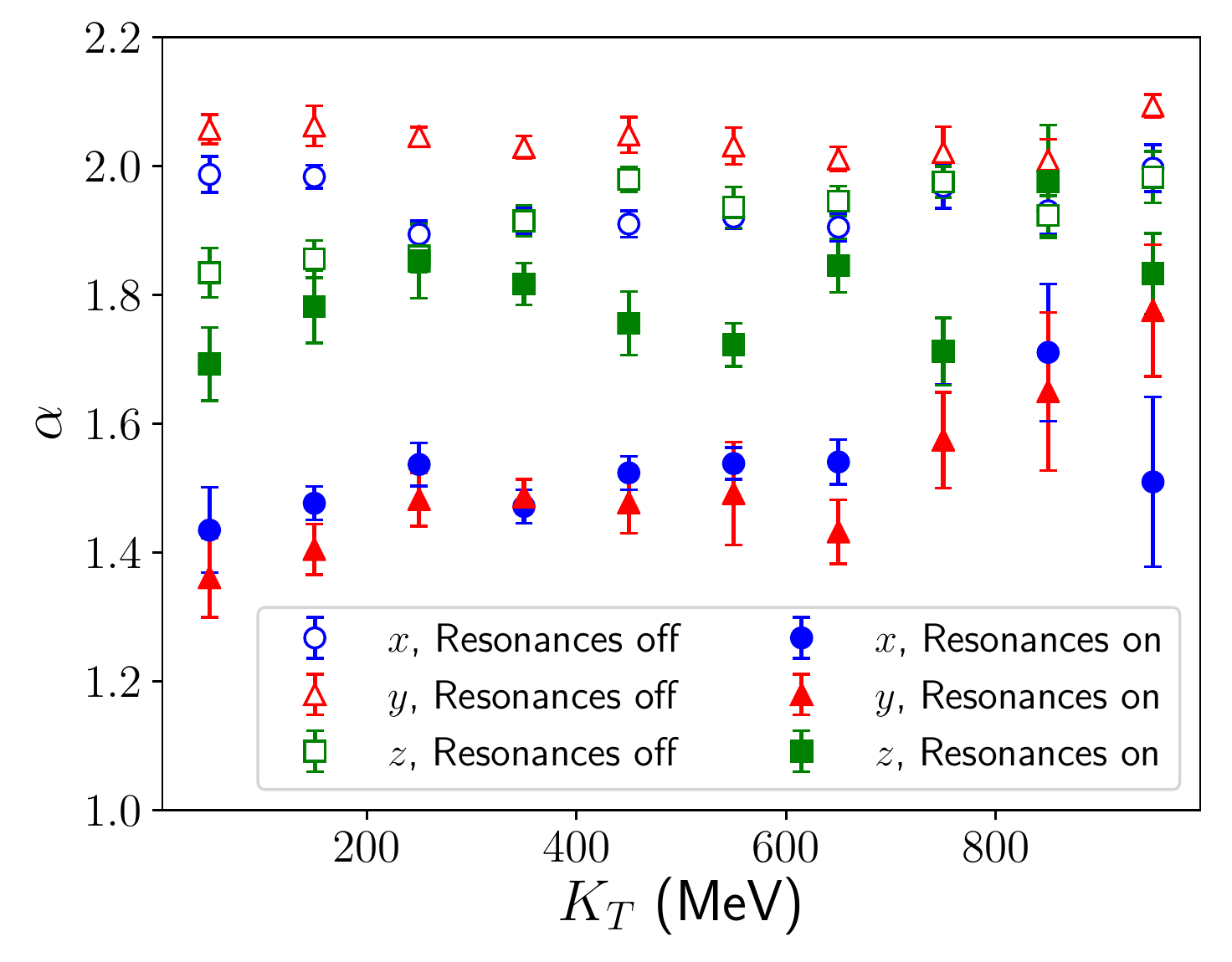}
\includegraphics[width=0.48\textwidth]{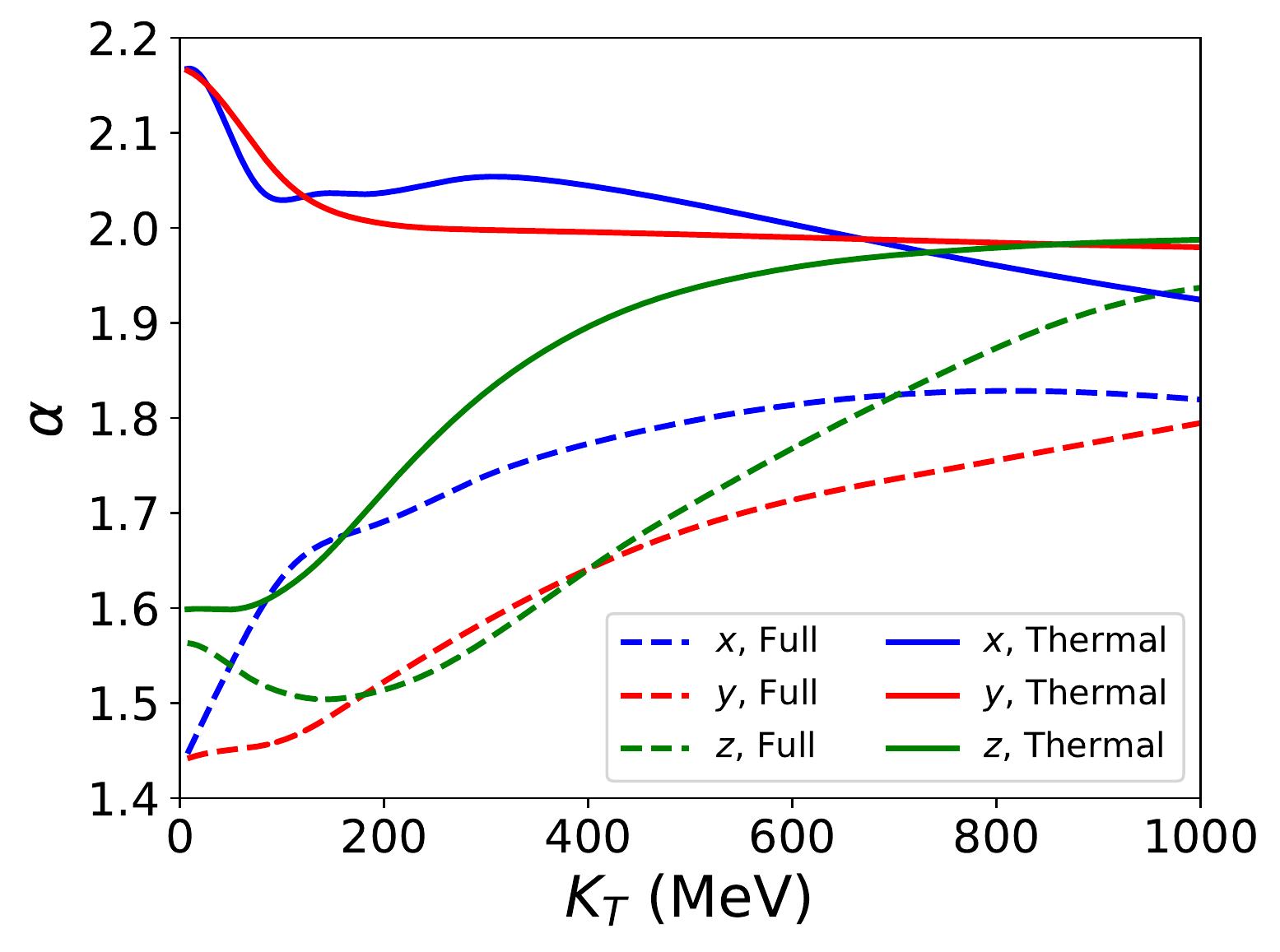}
\caption{The L\'evy index of the 1D fits to the correlation function in $\vec{q}$ along different axes, with or without resonances. Left panel: blast-wave model. Right panel: hydrodynamics.}
\label{fig4}
\end{figure}

To find out why does the 1D projection affect L\'evy index so significantly we have to look at the 3D correlation function. First, we fitted the correlation functions from both models in each direction separately. This is shown in Figure \ref{fig4}. These plots show, that while the correlation function behaves similar in outward and sideward direction, the $K_T$-dependence in longitudinal direction behaves differently. Moreover, it seems that the resonances do not affect the correlation function in the longitudinal direction as much as in the transverse plane. 

\begin{figure}[h]
\centering
\includegraphics[width=0.32\textwidth]{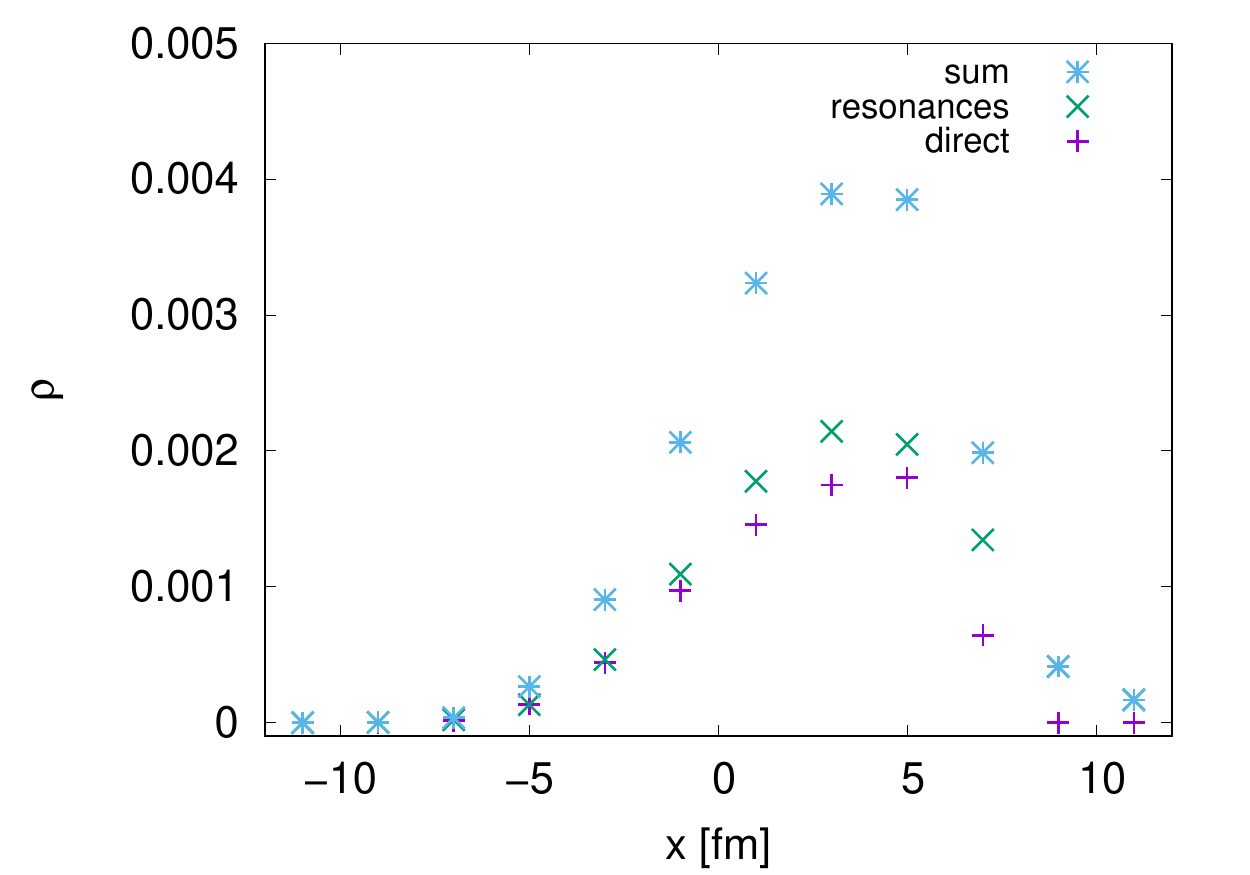}
\includegraphics[width=0.32\textwidth]{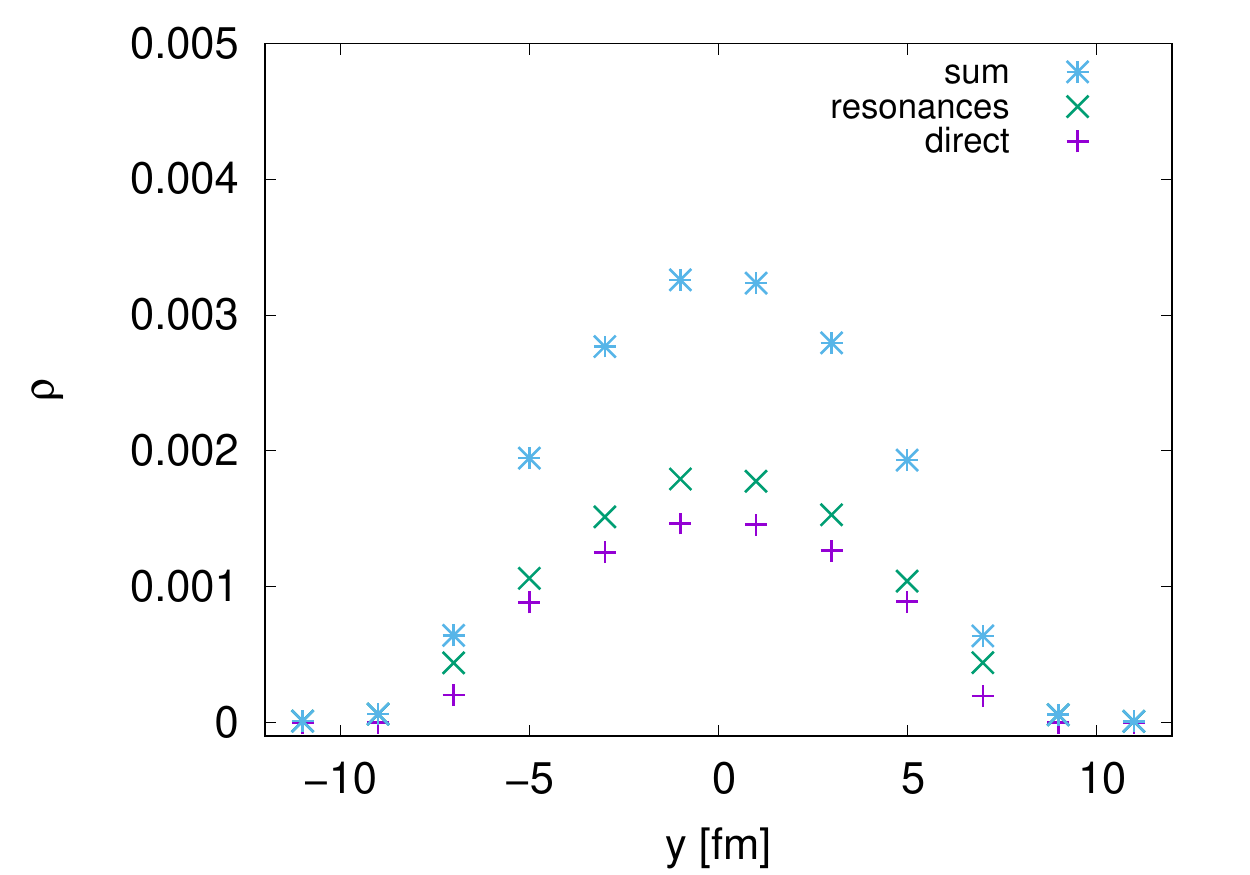}
\includegraphics[width=0.32\textwidth]{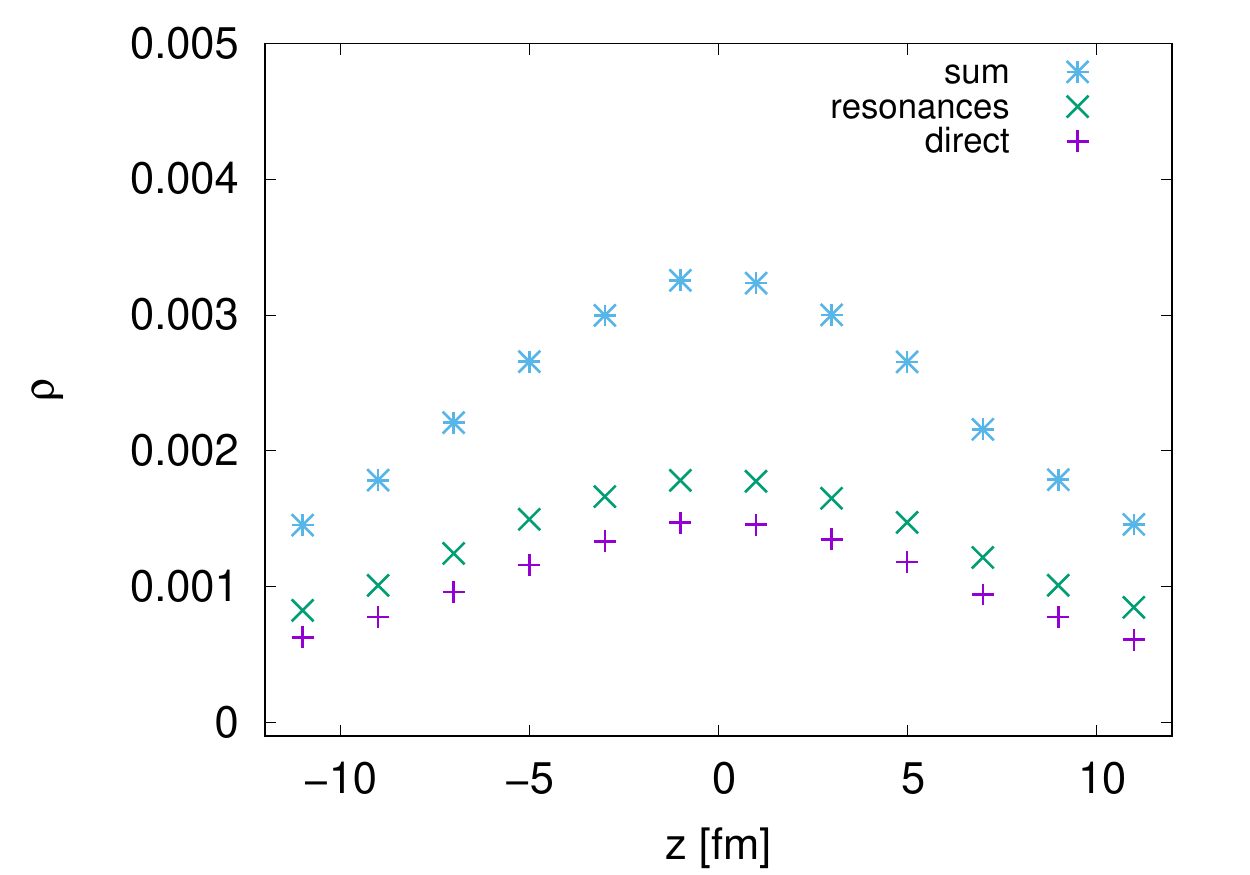}

\includegraphics[width=0.32\textwidth]{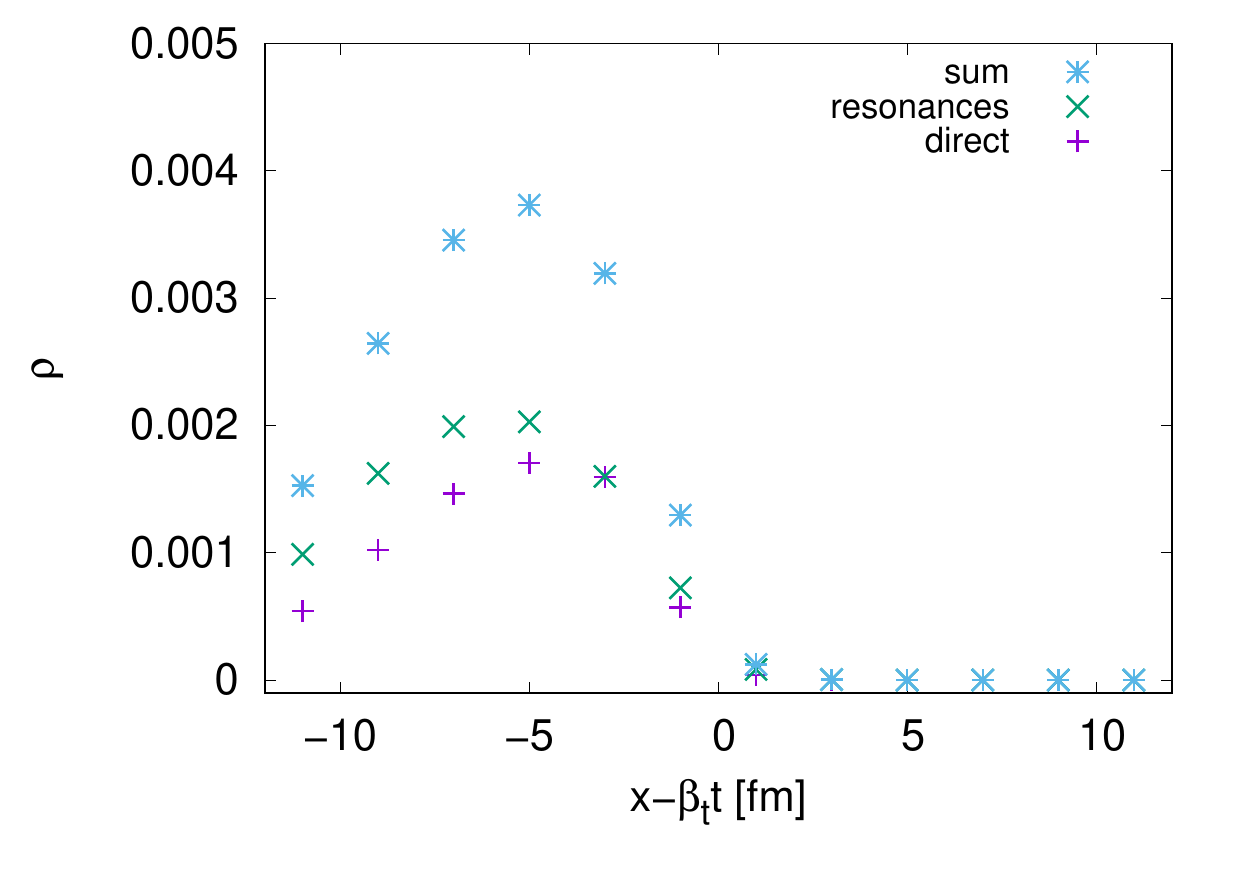}
\includegraphics[width=0.32\textwidth]{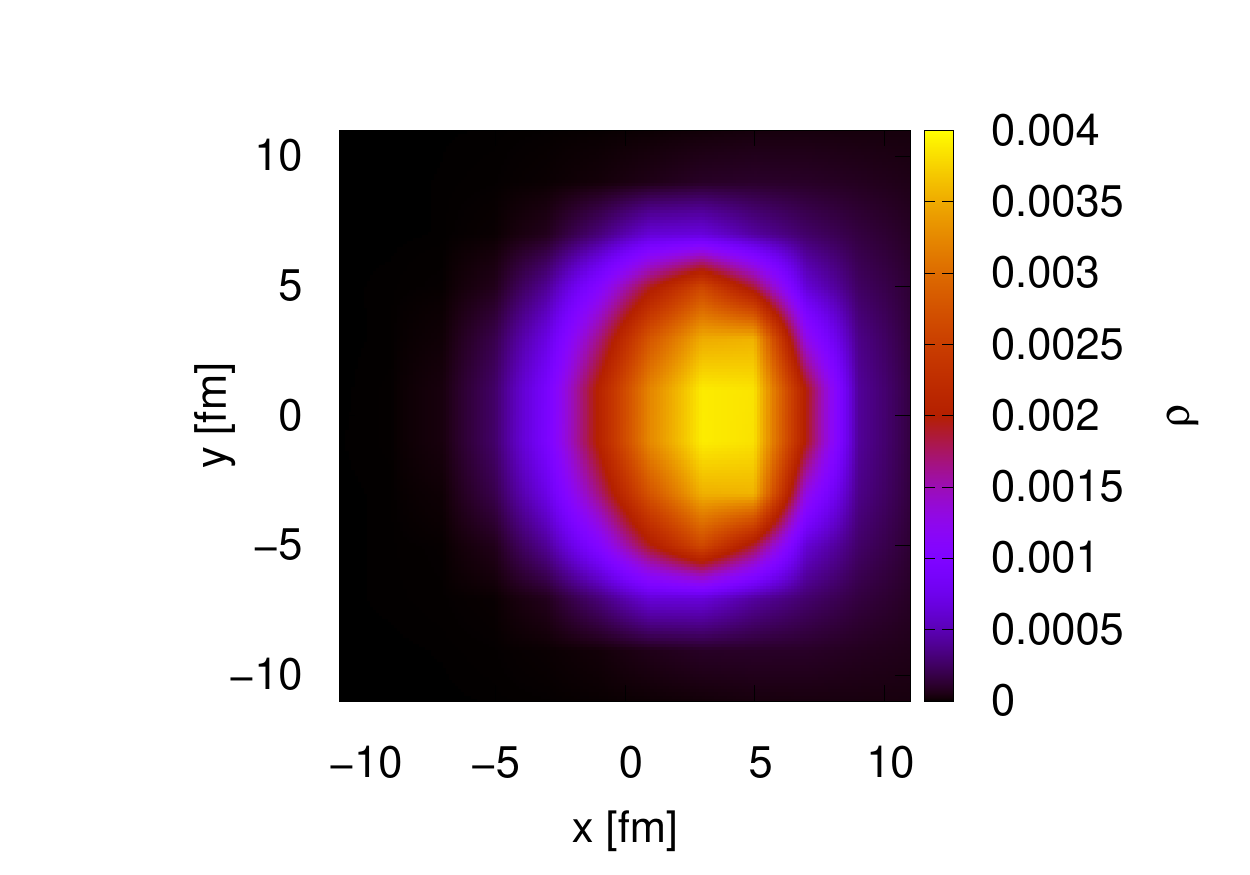}
\includegraphics[width=0.32\textwidth]{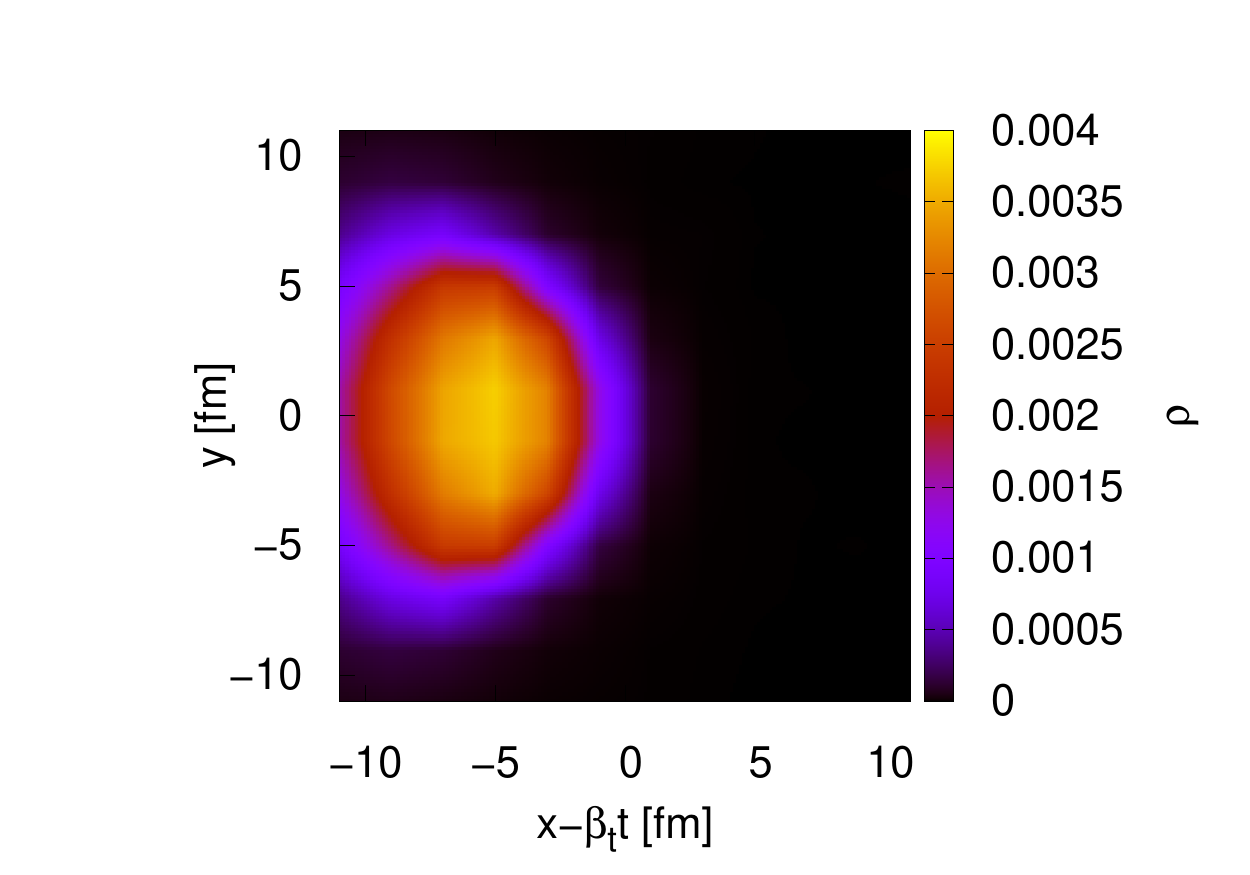}
\caption{The spatial distribution of the emission points of pions. Upper row: the profiles of the emission points distribution along the $x$ (left), $y$ (middle), and $z$-axis (right). Lower row: the profile along the variable $(x-\beta_t t)$ (left), and two-dimensional distributions in the transverse plane (middle and right). The green $\times$'s show the profile of direct pions, the blue $\star$'s show the profile of pions produced by resonances and purple $+$'s show their sum. All these distributions were calculated as narrow integrals over the remaining coordinates with width $2\:\unit{fm}$. }
\label{fig5}
\end{figure}

To illustrate why the behaviour in different directions is so different, we plotted spatial distributions of the emission points of pions using the blast-wave model. Figure \ref{fig5} shows us these profiles and we can see that even the source of pions along different axes looks unalike.

\begin{figure}[h]
\centering
\includegraphics[width=0.48\textwidth]{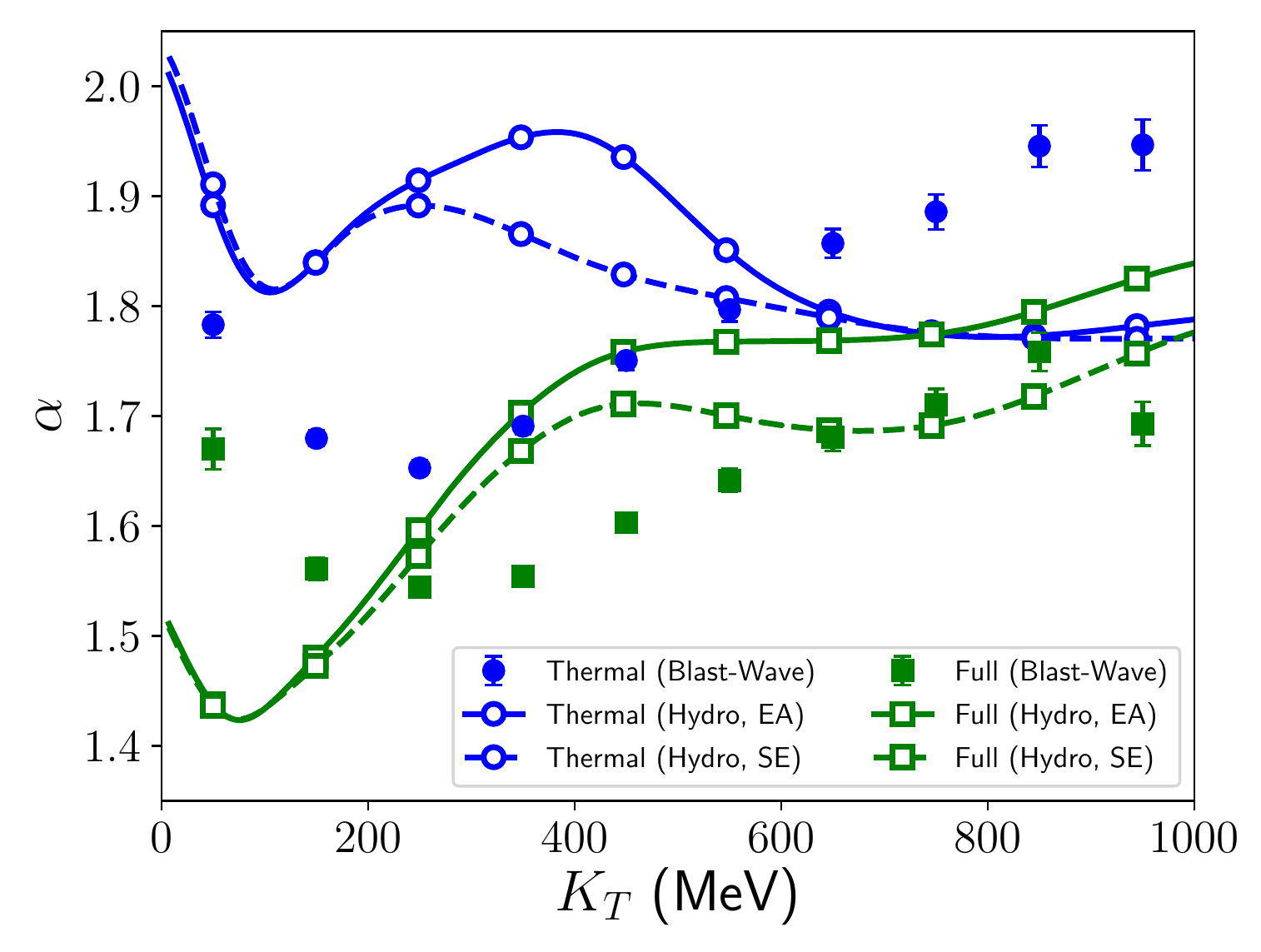}
\caption{The L\'evy index of the 3D fit to the correlation function according to Eq. \eqref{eq-3dlevy}, with and without resonances. The full points correspond to blast-wave model, while the empty points represent hydrodynamics.
Solid lines connect points for the event-averaged correlation functions, while dashed lines correspond to the correlation function for a single event.}
\label{fig6}
\end{figure}

To use the whole 3D correlation function for obtaining the L\'evy index, one can fit it with 3D L\'evy distribution using Eq. \eqref{eq-3dlevy}. 
Such parametrisation is then fit to the correlation functions in all bins of $q$, not just along the axes.
%Such fit then represents also parts of the correlation function we could not fit using 1D fits along axes. 
Figure \ref{fig6} shows the $K_T$-dependence of the L\'evy index obtained by a 3D fit of the correlation function. This figure underscores the fact that resonances can reduce the value of the L\'evy index independently of the chosen model.

%%%%%%%%%%%%%%%%%%%%%%%%%%%%%%%%%%%%%%%%%%%%%%%%%%%%%%%%%%%%%%%%%%%%%%%%%%%%

\section{Conclusions}

In this paper, we have shown that the shape of the correlation function, as well as the value of the L\'evy index, may be influenced by a variety of different mechanisms. All our results show that the L\'evy index may deviate substantially from the value of 2 due to non-critical effects. Some of these effects do not have significant influence, but others are found to cause notable deviations. The two most significant deviations arise, first, from the projection of the 3D relative momentum $\vec{q}$ onto a scalar $Q$, and second, from the inclusion of resonance decays. Since we used two different models, these results appear to be robust and not merely artifacts of the models we have used. For this reason, the conclusions presented here may be regarded as model-independent.

%%%%%%%%%%%%%%%%%%%%%%%%%%%%%%%%%%%%%%%%%%%%%%%%%%%%%%%%%%%%%%%%%%%

\subsubsection*{Acknowledgements}
This work was supported by the grant 17-04505S of the Czech Science Foundation (GA\v CR).
BT also acknowledges support from VEGA 1/0348/18 (Slovakia). CP is funded by the CLASH project (KAW 2017-0036) and gratefully acknowledges the use of computing resources from both the Minnesota Supercomputing Institute (MSI) at the University of Minnesota and the Ohio Supercomputer Center \cite{OSC} which contributed to the research results reported within this proceedings.

\end{document}